\documentclass[prl,twocolumn,showpacs]{revtex4}

\usepackage{amssymb}
\usepackage{graphicx}
\usepackage{amsmath}

\begin{document}

\title{Superconductivity near the vibrational mode instability in MgCNi$_{3}$%
.}
\author{A.~Ignatov, S. Y.~Savrasov, T.A.~Tyson}
\date{\today}

\begin{abstract}
To understand the role of electron-phonon interaction in superconducting
MgCNi$_{3}$ we have performed density functional based linear response
calculations of its lattice dynamical properties. A large coupling constant $%
\lambda $= 1.51 is predicted and contributing phonons are identified as
displacements of Ni atoms towards octahedral interstitials of the perovskite
lattice. Instabilities found for some vibrational modes emphasize the role
of anharmonic effects in resolving experimental controversies.
\end{abstract}

\maketitle

\address {Department of Physics, New Jersey Institute of Technology,
Newark, New Jersey 07102}

\draft

\input{epsf}

The discovery of superconductivity in MgCNi$_{3}$ \cite{Nature} has
generated a new puzzle in the recent series of found superconductors \cite%
{Recent}. Despite its relatively low T$_{c}$\symbol{126}8K, the presence of
Ni signals the possible importance of correlation effects which makes the
physics of the pairing mechanism relevant to the famous high T$_{c}$
cuprates and brings the discussion of unconventional non-electron-phonon
mechanism. The experimental information characterizes MgCNi$_{3}$ as
moderate \cite{Nature,Lin} or strong \cite{Mao} coupling conventional
superconductor through the analysis of specific--heat data, supports the
s--wave pairing by CMR experiments \cite{CMR}, and at the same time shows a
zero-bias anomaly in tunnelling data \cite{Mao}. A clear need for detailed
information about the phonon spectra emerges from these controversial data
in order to clarify the role of electron--phonon interaction (EPI)\ and
understand the mechanism of superconductivity.

In this work, we perform theoretical studies of the strength of the
electron--phonon coupling in MgCNi$_{3}$ by using fully self--consistent
density functional based linear response calculations \cite{PHNPRL} of the
lattice dynamical properties as a function of phonon wavevector $\mathbf{q}$%
. This method was proven to provide reliable estimates for the phonon
spectra and electron-phonon interactions in a large variety of systems\cite%
{PHNPRL,HTCPRL}. We extract a large coupling constant $\lambda =1.51$ and
identify phonons contributing to it as displacements of Ni atoms towards
octahedral interstitials of the perovskite--like cubic structure. We find
some of the lattice vibrations to be unstable in linear order which
emphasizes the role of strong anharmonic effects in the possible resolution
of the recent experimental puzzles.

The basic element of the perovskite MgCNi$_{3}$ structure is given by a
carbon atom placed at the center of the cube and octahedrally coordinated by
6 Ni atoms. Our electronic structure calculation using full potential linear
muffin-tin orbital (LMTO)\ method \cite{FPLMTO} reveals Ni-d C-p hybridized
valence bands $\epsilon _{\mathbf{k}j}$ in accord with the previous studies%
\cite{Mazin,Jarlborg,Shim}. The Fermi surface consists of several sheets
such as rounded cube sections centered at $\Gamma $, thin jungle gym area
spanning from $R$ $[\frac{1}{2}\frac{1}{2}\frac{1}{2}]\frac{2\pi }{a}$ to $M$
$[\frac{1}{2}\frac{1}{2}0]\frac{2\pi }{a}$points, dimpled square shaped hole
pockets centered around $X$ point $[\frac{1}{2}00]\frac{2\pi }{a}$ as well
as little ovoids along $\Gamma -R.$The tight--binding picture discussed
before \cite{Mazin} consists of carbon $p_{x}$ $p_{y}$ $p_{z}$ orbitals
hybridized with $d$ states of three Ni atoms numerated accordingly as Ni$%
_{x} $ Ni$_{y}$ and Ni$_{z}.$For example (see Fig. 1) carbon $p_{z}$ state
hybridizes with N\'{\i}$_{z}$ $d_{z^{2}-1^{2}}$ and also with N\'{\i}$_{x}$ $%
d_{xz}$ N\'{\i}$_{y}$ $d_{yz}$. Similar picture holds for carbon $%
p_{x},p_{y} $ orbitals which are not shown. As it has been noted \cite{Mazin}%
, this in particular results in two antibonding states crossing the Fermi
level, which in the nearest neighbor approximation have no dispersion along
some directions in the Brillouin zone (BZ).

\begin{figure}[tbh]
\includegraphics*[height=2.5in]{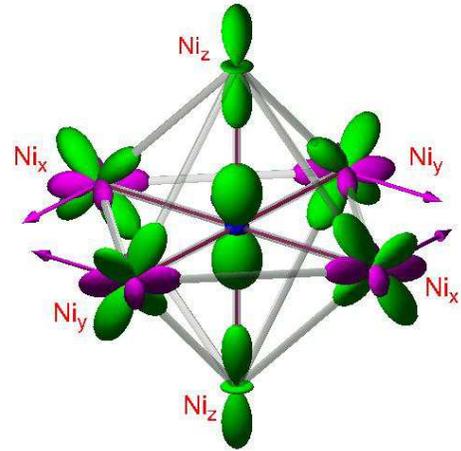}
\caption{Basic element of the structure and set of tight-binding orbitals
relevant to low-energy physics MgCNi$_{3}$. Arrows show displacements of Ni
atoms corresponding to wavevector $\mathbf{q}=[\frac{1}{2}\frac{1}{2}0]\frac{%
2\protect\pi }{a}$.}
\label{fig:Fig1}
\end{figure}

The appearance of nearly flat areas of $\epsilon _{\mathbf{k}j}$ gives rise
to 2D van Hove singularity (vHS) placed 40 meV below the Fermi energy $E_{F}$
which is responsible for a strong narrow peak near $\epsilon _{F}$. This has
generated speculation about closeness of MgCNi$_{3}$to ferromagnetic
instability upon doping \cite{Mazin,Jarlborg,Shim,Pickett}. The narrowness
of the vHS band is controlled by the second--nearest neighbor hopping
integrals, which for the states shown in Fig.1 correspond to the overlap
between Ni$_{x}$ and Ni$_{y}$ $d_{xy}$ orbitals. As we discuss in this work,
nearly unstable phonon modes exists when each of the two Ni atoms moves
toward octahedral interstitial sites. (This is shown in Fig. 1 by arrows for
a phonon wave vector $\mathbf{q}=[\frac{1}{2}\frac{1}{2}0]\frac{2\pi }{a}$)$%
. $ We shall see that these distortions wipe out the narrow vHS peak and
give rise to a large electron--phonon coupling.

To calculate lattice dynamics of MgCNi$_{3}$ as a function of wave vector $%
\mathbf{q}$ we utilize the linear response method\cite{PHNPRL,FPLMTO}, We
use $2\kappa $ LMTO\ basis set, generalized gradient approximation for
exchange--correlation\cite{GGA}, experimental lattice constant a=7.206 a.u.,
as well as effective (40,40,40) grid in $\mathbf{k}$-space (total 1771
irreducible $\mathbf{k}$ points) to generate the phonon dispersions $\omega
_{\mathbf{q}\nu }$ and electron--phonon matrix elements $g_{\mathbf{k}+%
\mathbf{q}j^{\prime }\mathbf{k}j}$on a (10,10,10) grid of the $\mathbf{q}$
vectors (total 56 irreducible $\mathbf{q}$ points).

Our calculated phonon spectrum along major high symmetry lines of the cubic
Brillouin zone is given on Fig. 2. The frequencies are seen to be span up to
900\ K, with some of the modes showing significant dispersion. In general,
we distinguish three panels where the top three branches around 900 K are
carbon based, the middle three branches around 600 K are Mg based and 9
lower branches are all Ni based. For the $\Gamma $ point z--polarized modes,
in particular, consist of (i) Ni$_{z}$-C against Mg-Ni$_{x}$-Ni$_{y}$
vibrations (186 K), (ii)\ pure Ni$_{x}$-Ni$_{y}$ vibrations (262 K), (iii)
Mg against Ni$_{x}$-Ni$_{y}$\ (561\ K) (iv) C against Ni$_{z}$
vibrations(857 K).

\begin{figure}[tbh]
\includegraphics*[height=2.8in]{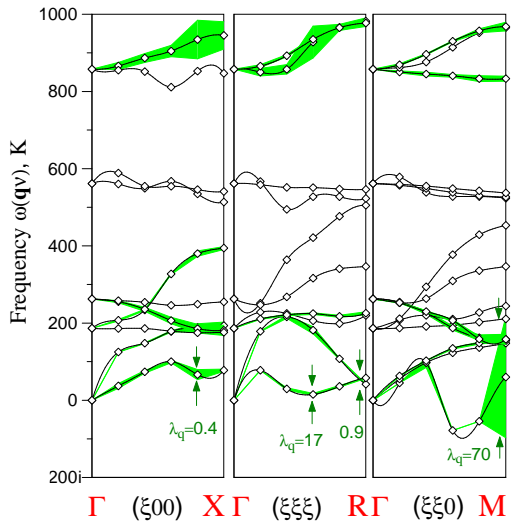}
\caption{Calculated phonon spectrum of MgCNi$_{3}$ using density functional
linear response method. Some curves are widened proportionally to the phonon
linewidths.}
\label{fig:Fig2}
\end{figure}

A striking feature of this phonon spectrum is the presence of a
low-frequency acoustic mode which is very soft and is even seen to be
unstable along $(\xi \xi 0$) direction in the BZ. This mode is essentially
Ni based and corresponds to perpendicular movements of two Ni atoms towards
octahedral interstitials of the perovskite structure. The latter is made of
each of the four Ni atoms and two Mg atoms. For example, considering the xy
plane (see Fig. 1) for the $\mathbf{q}$ point M such movements can be seen
as a 2D breathing around this vacant interstitial. We find a similar
situation for other wave vectors and in other directions of the BZ, where
each pair of Ni atoms prefers such in-phase displacements perpendicular to
each other. The softness and instability here can be understood as the
octahedral interstitials are only places to escape for each Ni atom stressed
between two surrounding carbons.

The discussed displacements affect the overlap integrals between nearest Ni t%
$_{2g}$ orbitals which control the width of the vHS band. For the xy plane
these are the hoppings between $d_{xy}$ orbitals (see Fig. 1). It is
therefore clear that these modes should have large EPI. As they bend the
Ni--C bonds, this effect has already been previously emphasized\cite{Mazin}
by considering rotations of the Ni based octahedra at the R point. Here, we
point out movements for wave vectors away from R point, where, for example,
Ni$_{x}$--C and Ni$_{y}$--C bonds can be seen as changing their relative
angle.

\begin{figure}[tbh]
\includegraphics*[height=2.5in]{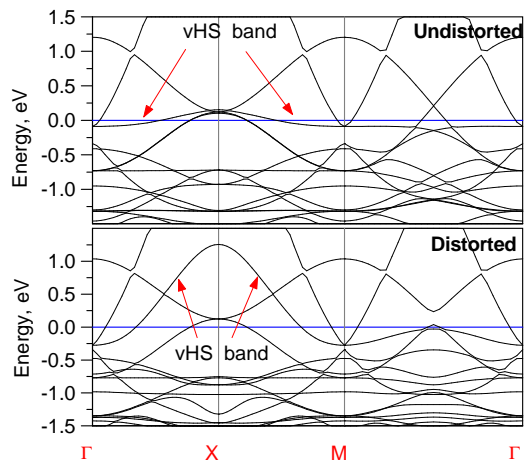}
\caption{Calculated one--electron structure corresponding to the Ni based
frozen phonon with $\mathbf{q}=[\frac{1}{2}\frac{1}{2}0]\frac{2\protect\pi }{%
a}$. Top panel - undistorted bands, bottom panel - distroted bands
corresponding to the Ni displacements by 0.2 \AA .}
\label{fig:Fig3}
\end{figure}

To illustrate the crucial change in the electronic structure due to such
distortions, Fig. 3 shows two one--electron spectra in the vicinity of the
Fermi level corresponding to the M point frozen phonon involving N\'{\i}$%
_{x} $ displacement along the y axis in phase with the Ni$_{y}$ displacement
along the x axis as illustrated on Fig. 1. The top panel of Fig. 3
corresponds to the undistorted energy bands drawn in the original cubic BZ
for easier comparison with the published data \cite{Mazin,Jarlborg,Shim}. As
the point $[\frac{1}{2}\frac{1}{2}0]\frac{2\pi }{a}$ is now a reciprocal
vector of the new doubled lattice, the bands are seen to be simply folded,
and the narrow vHS band is readily recognized. When we introduce a
distortion by 0.2 \AA , the only essential difference is the width of the
vHS band which now disperses as much as 1.5 eV. Given the smallness of the
assumed displacement this emphasizes the large electron--phonon coupling.
Its appearance cannot be understood from a simple geometric overlap between
the two $t_{2g}$ orbitals ($d_{xy}$ states between Ni$_{x}$ and Ni$_{y}$
shown on Fig. 1). We therefore look for an electronic enhancement due to
nesting--like features of the Fermi surface. Indeed, such nesting can be
found for the two dimpled square shaped hole pockets centered at X separated
exactly by the wave vector $[\frac{1}{2}\frac{1}{2}0]\frac{2\pi }{a}.$ We
have confirmed that feature by corresponding calculation of the integral $%
\sum_{\mathbf{k}jj^{\prime }}\delta (\epsilon _{\mathbf{k}j}-\epsilon
_{F})\delta (\epsilon _{\mathbf{k+q}j}-\epsilon _{F})$ which provides the
total phase space available for the electrons to scatter at given wave
vector $\mathbf{q}$ with no energy change.

We now turn our discussion to the detailed dependence of the
electron--phonon coupling across the entire BZ. This at least can be done
for all stable phonons. Fig.2 shows the calculated phonon linewidths $\gamma
_{\mathbf{q}\nu }$by widening some representative dispersion curves $\omega
_{\mathbf{q}\nu }$ proportionally to $\gamma _{\mathbf{q}\nu }$. Each phonon
linewidth is proportional to \cite{Allen72} $\ \sum_{\mathbf{k}jj^{\prime
}}|g_{\mathbf{k}+\mathbf{q}j^{\prime }\mathbf{k}j}^{\mathbf{q}\nu
}|^{2}\delta (\epsilon _{\mathbf{k}j}-\epsilon _{F})\delta (\epsilon _{%
\mathbf{k+q}j}-\epsilon _{F})$ where the electron--phonon matrix element is
found self--consistently from the linear response theory \cite{PHNPRL}. In
particular, we see that some phonons have rather large linewidhts. This, for
example holds, for all carbon based higher lying vibrational modes. The
strength of the coupling, $\lambda _{\mathbf{q}\nu },$ for each mode can be
obtained by dividing $\gamma _{\mathbf{q}\nu }$ by $\pi N(\epsilon
_{F})\omega _{\mathbf{q}\nu }^{2},$where $N(\epsilon _{F})$ is the density
of states at the Fermi level equal to 5.3 st./[eV*cell] in our calculation.
Due to large $\omega _{\mathbf{q}\nu }^{2},$ this unfortunately results in
strongly suppressed coupling for all carbon modes which would favor high
critical temperatures. The coupling, however, is relatively strong for the
Ni based modes. For example, we can find $\lambda $'s of the order of 1--3
for the Ni based optical phonons around 250\ K. Again, the analysis of the
polarization vectors shows that these vibrations involve Ni movements
towards octahedral interstitials. For example, (see Fig. 1) the movement of
Ni$_{x}$ and Ni$_{y}$ atoms along the z direction either in-phase or out-of
phase result in larger overlap between Ni$_{z}$ Ni$_{y}$ $d_{yz}$ orbitals
and between Ni$_{z}$ Ni$_{x}$ $d_{xz}$ orbitals. Similar to what we find for
the M point using the frozen phonon method (Fig.3), this again enlarges the
vHS bandwidth resulting in large EPI. A extremely large coupling ($\lambda 
\symbol{126}70)$ occurs for the soft acoustic mode involving the
interstitial breathing (M--point). Here we point out a triple effect: (i)
breathing of four Ni$_{x}$ Ni$_{y}$ atoms into the interstitial, which
results in larger $d_{xy}$ overlap, (ii) nesting enhancement which helps
wiping out the vHS peak, and, (iii) the smallness of $\omega _{\mathbf{q}\nu
}^{2}.$As the linewidth of this particular phonon is so large, the concept
of phonon itself has to be questioned, but due to a smallness of the phase
space associated with this vibration, this has a little effect on integral
characteristics such as $\lambda .$

Unfortunately, finding the integral value of $\lambda $ is another
challenging problem due to the appearance of the imaginary frequencies.
Neglecting the unstable mode completely results in the average coupling
constant equal to 0.95 mainly due to the discussed Ni vibrations around 250\
K (see Fig. 2).. It is however clear that the low--frequency mode has a
large contribution to $\lambda $ and cannot be omitted. We have asked a
question for which wave vectors $\mathbf{q}$ this mode is unstable? Since we
know its dispersion across the entire BZ, we can determine a surface in $%
\mathbf{q}$ space which separates the real and imaginary frequencies. Fig. 4
shows the result of such an analysis. We see the area around the $\Gamma $
point which continues along the lines towards the $R,M,$ and $X$ points.
Here we find the stability of the mode. The area around the M point, where
we find enormously large EPI, is seen to be very small . This is also clear
from the dispersion relations shown on Fig. 2 for the ($\xi \xi 0$)
direction where the frequency becomes real just near the M point itself.

\begin{figure}[tbh]
\includegraphics*[height=2.2in]{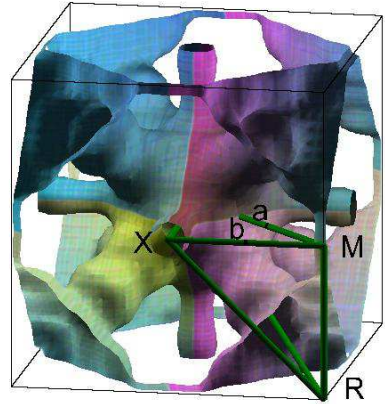}
\caption{Surface in the Brillouin zone which separates stable and instable
areas for the Ni based acoustic mode.}
\label{fig:Fig4}
\end{figure}

This instability carefully avoids symmetry points which can be understood by
keeping in mind the symmetry of the discussed distortions. Namely,
anharmonicity is expected to be large for all in--plane movements involving
two, three, or four Ni atoms towards octahedral interstitials. This is
forbidden near the $\Gamma $ point and for all wavevectors $(\xi 00)$ along $%
\Gamma X$ as well as for $(\xi \xi \xi )$ along $\Gamma R$ directions if,
for example, xy plane is considered. It is however allowed for $(\xi \xi 0)$
along $\Gamma M\ $and for $(\frac{1}{2}\xi 0)$ along $XM.$ Moreover, for the
wavevectors with $q_{z}\neq 0,$ the displacements away from xy plane are
allowed and the instability is quickly suppressed. This, \textit{e.g.}, is
the case of $(\frac{1}{2}\frac{1}{2}\xi )$ along $MR.$

The persistence of the instability which does not occur for any of the
high--symmetry point needs a non--trivial frozen--phonon analysis. As our
polarization vectors promt that the largest anharmonicity is expected for
the $\Gamma XM$ plane with $q_{z}=0$, we have performed three such
calculations for the points $a=$($\frac{1}{4}\frac{1}{4}0)\frac{2\pi }{a}$, $%
b=$($\frac{1}{2}\frac{1}{4}0)\frac{2\pi }{a}$, and $M$ shown on Fig. 4. This
corresponds to the discussed displacements of two (point $a$), three (point $%
b$), or four (point $M$) Ni atoms within xy--plane. The results of these
calculations reveal esentially anharmonic interatomic Ni potentials. A
shallow double well with a depth of the order of 20--40 K and the curvature
at the equillbrium of the order of 40--50i K exists for the $a$ and $b$
points which becomes vanishing at the M point. We have found such a behavior
by both total energy and force calculations for the supercells of 20 atoms ($%
a$ and $b$ points) and 10 atoms ($M$ point). Clearly, such a small depth on
the temperature scale indicates that the distortions are dynamical and zero
point motions would prevent the appearance of the static long--range order.
This picture is confirmed by our recent extended x--ray absorbtion fine
spectroscopy experiments (EXAFS) which reveal dynamical distortions \cite%
{Ignatov} not seen in the diffraction experiments. Remarkably that the
parameters of the double well (depth \symbol{126}20K and the equillbrium
curvature \symbol{126}50i K) extracted from EXAFS are in accord with our
frozen phonon data.

It is now clear that our instabile mode is not related to a statically
distorted structure of MgCNi$_{3}$ but should be resolved using an
anharmonic theory of the EPI. Using a combination of our frozen phonon and
linear response data we are able to map the behavior of the instable mode
inside the entire Brillouin zone using a (4,4,4) grid of $\mathbf{q}$%
--vectors. This gives us a possibility to determine a contribution to the
Eliashberg spectral function using a generalized formula \cite{Allen,BKBO}%
\begin{eqnarray}
&&\alpha ^{2}F(\omega )=\frac{1}{N(\epsilon _{F})}\sum_{\mathbf{q}}\sum_{%
\mathbf{k}j\mathbf{k}^{\prime }j^{\prime }}\sum_{n}\delta \lbrack \omega
-\omega _{n}(\mathbf{q})]\times   \notag \\
&&\frac{[f_{\mathbf{k}j}-f_{\mathbf{k}^{\prime }j^{\prime }}]\delta \lbrack
\epsilon _{\mathbf{k}j}-\epsilon _{\mathbf{k}^{\prime }j}+\omega _{n}(%
\mathbf{q})]|G_{\mathbf{k}^{\prime }j^{\prime }\mathbf{k}j}^{(n)}(\mathbf{q}%
)|^{2}}{[\omega _{n}(\mathbf{q})]^{2}}  \label{Anharmonic}
\end{eqnarray}%
where the summation over the eigenstates $n$ of the anharmonic oscillator
appears here with $\omega _{n}(\mathbf{q})=\epsilon _{n}(\mathbf{q}%
)-\epsilon _{0}(\mathbf{q})$ being the excitation frequencies around the
ground state level for the anharmonic well at a given wave vector $\mathbf{q}
$. The generalized electron--phonon matrix element $G_{\mathbf{k}^{\prime
}j^{\prime }\mathbf{k}j}^{(n)}(\mathbf{q})$ involves transitions from the
ground to $n$th excited state and also includes changes in the effective
one--electron potential to all orders with respect to the displacements. The
most important is the effect of the frequency renormalization which appears
in denominator of Eq. (\ref{Anharmonic}). As we find, our anharmonic mode
has the first excitation frequency which, depending on the $\mathbf{q}$%
--point, varies from 120 to 150\ K. We also find that the summation over $n$
is fastly convergent in accord with the previous work\cite{Allen}: the
oscillator strength for $n$=2 is always less than 2\% of that for $n$=1. The
linear electron--phonon scattering matrix elements are known to us from the
linear response calculations, and numerically small corrections due to
second--order couplings have been extracted from our frozen phonon data
using the band splittings technique \cite{BKBO}.

Our resulting value of $\lambda $ for this anharmonic mode appears to be
0.56. Adding the result for $\lambda =0.95$ from all other modes our total $%
\lambda $ is now 1.51. This is consistent with the values of $\lambda $
extracted from specific heat measurements \cite{Nature,Lin,Walte} and would
cause MgCNi$_{3}$ to be a strongly coupled electron--phonon superconductor.
To estimate the T$_{c},$ we solve the Eliashberg equation with our total
(harmonic+anharmonic) $\alpha ^{2}F(\omega ).$ The cutoff frequency $\omega
_{cut}$ is taken to be 10 times the maximum phonon frequency. The Coulomb
pseudopotential $\mu ^{\ast }(\omega _{cut})$ is varied and the T$_{c}$
within the range 7--20 K is obtained. To make connections with the values of 
$\mu ^{\ast }$ which are standardly discussed with the Allen--Dynes modified
McMillan T$_{c}$ expression \cite{McMillan} we need to use a rescaling
formula \cite{V} $[\mu ^{\ast }]^{-1}=[\mu ^{\ast }(\omega _{cut})]^{-1}+\ln
(\omega _{cut}/\omega _{\log })$ with $\omega _{\log }\sim 120K.$ We found
that that both the experimental $T_{c}$ and the superconducting energy gap $%
\Delta (0)$ can be reproduced with the value of $\mu ^{\ast }=0.33$. While
the effect of spin fluctuations on superconductinvity and mass enchancement
is better to discuss in terms of its own coupling constant $\lambda _{spin}$%
, at the absence of detailed calculation of the latter, we can only conclude
that our enchanced $\mu ^{\ast }$ can in principle appear due to localized
nature of Ni $d$ orbitals. It is, for example, known that spin fluctuatlions
are important in V and $\mu ^{\ast }\symbol{126}0.3$ is needed\cite{V}$.$ In
fact, at the absence of spin fluctuations, $\mu ^{\ast }$ is usually
0.1--0.15 and T$_{c}$ in MgCNi$_{3}$would be larger. We conclude that spin
fluctuations partially suppress superconductivity, the result expected from
the conventional theory. Note that the same conclusion has been reached
based on the recent analysis of the specific heat data \cite{Walte}. We
finally calculated the phonon contribution to the specific heat and fitted
it to the form $C(T)=\beta T^{3}$ at low temperatures. Our value of $\beta
=0.35$ mJ/[mol*K$^{4}$] is close to the values 0.40--0.42 mJ/[mol*K$^{4}$]
determined experimentally \cite{Lin,Mao,Walte}.

In conclusion, by performing linear response calculations of the
electron--phonon interaction in MgCNi$_{3}$ we reported the value of $%
\lambda =1.51$ consistent with the strong--coupling limit of
electron--phonon mechanism of superconductivity. The unusually large
anharmonic correction to $\lambda $ for the lattice near instability is
emphasized.

We thank O. V. Dolgov, I. I. Mazin, W. Pickett for useful comments, and S.
Shulga for availability of his programs to solve Eliashberg's equations. The
work was supported by the grants NSF DMR No.9733862, 0209243,  0238188,
0342290, US\ DOE No. DE--FG02--99ER45761, NJSGC No. 02--42.


\begin{references}
\bibitem{Nature} T. He {\it et.al}, Nature {\bf 411}, 54 (2001).

\bibitem{Recent} J. Nagamatsu {\it et. al}., Nature {\bf 410}, 63 (2001); S.
Uji {\it et.al}, Nature {\bf 410}, 908 (2001); S. S. Saxena {\it et.al},
Nature {\bf 406}, 587 (2000); Matzdorf {\it et. al}, Science {\bf 289}, 746
(2000).

\bibitem{Lin} J.-Y. Lin, {\it et.al.}, Phys. Rev. B {\bf 67}, 052501 (2003).

\bibitem{Mao} Z. Q. Mao {\it et.al.}, Phys. Rev. B {\bf 67}, 094502 (2003).

\bibitem{CMR} P. M. Singer, {\it et.al,} Phys.\ Rev. Lett. {\bf 87}, 257601
(2001)

\bibitem{PHNPRL} S.Y.~Savrasov {\it et.al.}, Phys. Rev. Lett. {\bf 69}, 2819
(1992); {\it ibid}, {\bf 72}, 372 (1994).

\bibitem{HTCPRL} S.Y. Savrasov, {\it et.al.}, Phys. Rev. Lett. {\bf 77},
4430 (1996);{\it \ ibid.} {\bf 90}, 056401 (2003).

\bibitem{FPLMTO} S. Y. Savrasov, Phys. Rev. B {\bf 54}, 16470 (1996).

\bibitem{Mazin} D. J. Singh {\it et.al.}, Phys. Rev. B {\bf 64}, 140507(R)
(2001).

\bibitem{Jarlborg} S. B. Dugdale {\it et.al.}, Phys. Rev. B {\bf 64},
100508(R) (2001).

\bibitem{Shim} J. H. Shim {\it et.al.}, Phys. Rev. B {\bf 64}, 180510(R)
(2001).

\bibitem{Pickett} H. Rosner {\it et.al.}, Phys. Rev. Lett. {\bf 88}, 027001
(2002).

\bibitem{GGA} J. P. Perdew {\it et.al.}, Phys. Rev. Lett. {\bf 77}, 3865
(1996).

\bibitem{Allen72} P. B. Allen, Phys. Rev. B {\bf 6}, 2577 (1972).

\bibitem{Ignatov} A. Ignatov {\it et.al}, Phys. Rev. B {\bf 67}, 064509 (2003).

\bibitem{Allen} J. C. K. Hui, P. B. Allen, J. Phys. F {\bf 4}, L42 (1974).

\bibitem{BKBO} V. Meregalli, {\it et.al }Phys. Rev. B {\bf 57}, 14453 (1998).

\bibitem{Walte} A. W\"{a}lte {\it et. al}, cond--mat/0208364.

\bibitem{McMillan} P. B. Allen and R. C. Dynes, Phys. Rev. B {\bf 12}, 905
(1975).

\bibitem{V} S. Y. Savrasov, {\it et.al.} Phys. Rev. B {\bf 54}, 16487 (1996).

\bibitem{RHO} T. G. Kumary, Phys. Rev. B {\bf 66}, 064510 (2002).
\end{references}
\end{document}